\documentclass{article}
\pdfoutput=1

\usepackage[nonatbib, preprint]{neurips_2021}

\usepackage[utf8]{inputenc} % allow utf-8 input
\usepackage[T1]{fontenc}    % use 8-bit T1 fonts
\usepackage{hyperref}       % hyperlinks
\usepackage{url}            % simple URL typesetting
\usepackage{booktabs}       % professional-quality tables
\usepackage{amsfonts}       % blackboard math symbols
\usepackage{nicefrac}       % compact symbols for 1/2, etc.
\usepackage{microtype}      % microtypography
\usepackage{xcolor}

\usepackage{amsmath}
\usepackage{amsfonts}       % blackboard math symbols
\usepackage{amssymb, bm}
\usepackage{pifont}% http://ctan.org/pkg/pifont
\newcommand{\cmark}{\ding{51}}%
\newcommand{\vi}{\mathbf{v}_i}
\newcommand{\xmark}{--}%

\usepackage{graphicx}
\usepackage{wrapfig}

\newcommand{\bt}{\bm{\theta}}
\newcommand{\bp}{\bm{\phi}}

\title{Relational VAE: A Continuous Latent Variable Model for Graph Structured Data}

\author{
  Mylonas Charilaos \\
  Structural Mechanics and Monitoring\\
  ETH Zurich\\
  \texttt{mylonas.charilaos@gmail.com} \\
   \And
   Imad Abdallah\\
   Structural Mechanics and Monitoring\\
   ETH Z\"urich\\
   \texttt{abdallah@ibk.dbaug.ethz.ch}\\
   \And
  Eleni N. Chatzi \\
  Structural Mechanics and Monitoring\\
  ETH Zurich\\
  \texttt{chatzi@ibk.dbaug.ethz.ch} \\
}

\begin{document}
\maketitle

\begin{abstract}
Graph Networks (GNs) enable the fusion of prior knowledge and relational reasoning with flexible function approximations. In this work, a general GN-based model is proposed which takes full advantage of the relational modeling capabilities of GNs and extends these to probabilistic modeling with Variational Bayes (VB). To that end, we combine complementary pre-existing approaches on VB for graph data and propose an approach that relies on graph-structured latent and conditioning variables. It is demonstrated that Neural Processes can also be viewed through the lens of the proposed model. We show applications on the problem of structured probability density modeling for simulated and real wind farm monitoring data, as well as on the meta-learning of simulated Gaussian Process data. We release the source code, along with the simulated datasets.
\end{abstract}

\newcommand{\G}{{G}}
\newcommand{\V}{\mathcal{V}}
\newcommand{\E}{\mathcal{E}}
\newcommand{\ie}{i.e., }

\section{Introduction}

Graph Neural Networks (GNNs) \cite{gilmer2017neural, graphnetpaper}
have been established as an effective tool for representation learning 
on graph structured data. 
Graph structured data are routinely employed to represent entities and relations
among them. The present work focuses in representation of uncertainty and generative
modeling for attributed directed graph data with continuous attributes.
The initiating motivation for this work is the ubiquity of noisy structured data and systems with stochastic or partially observable interactions of industrial relevance (e.g. wind farms and urban transportation networks).

In the context considered herein, modeled entities (\textit{nodes}) and 
modeled relations (\textit{edges}) may feature a \textit{state}, which may not 
be fully observed and/or stochastic. The same may also hold for
global (\textit{graph}) attributes. At the same time, nodes and relations may 
possess a dynamic partially observed state, which we may infer directly from 
data. Both the node states and edge states are not fully observed and non-deterministic, 
which amply motivates probabilistic extensions of graph networks. In essence, 
this work proposes a method that 1) exploits the relational structure of 
data and 2) allows for learning flexible distributions over entity and 
relation attributes.
Several partially overlapping approaches for this problem exist. A short review of
such prior approaches is offered in \autoref{sec:gnnreview}. Modeling entities and relations has been shown empirically to allow for stronger generalization \cite{pmlr-v119-sanchez-gonzalez20a,pfaff2020learning,zambaldi2018relational} in novel settings.
The main contribution of this work is to propose an approach to transfer the potent combinatorial generalization and modeling capabilities of GNNs to the problem of modeling conditional distributions of structured data.

\section{Methods}
\paragraph{Attributed graphs} Following \cite{graphnetpaper}, global attribute augmented graphs are denoted by $G = (\mathcal{V},\mathcal{E},\mathbf{u})$ where $\mathcal{V}:\{ \mathbf{v}_i\}_{i=1:N^v}$ with $\mathbf{v}_{i} \in \mathbb{R}^{d^v}$ denoting the nodes (vertices) of the graph, $\mathcal{E} :\{ (\mathbf{e}_{k}, s_{k}, r_{k}) \}_{k=1:N^e}$ designating the set of edges, with edge attributes $\mathbf{e}_k \in \mathbb{R}^{d^{e}} $, $s_k, r_i \in \mathbb{N}^1$ denote the head (sender) and tail (receiver) nodes of the modeled relation, while $\mathbf{u} \in \mathbb{R}^{d^u}$ is the global attribute.

\paragraph{Graph Networks (GN)} (or GraphNets) are composite functions that receive and return attributed graphs. The full GN block consists of an edge update, a node update and a global update block. Each block contains a corresponding function $\phi^{e},\phi^v, \phi^u$. The edge update function uses edge, node and global data. The edge block is followed by an aggregation step $\rho^{e \rightarrow n}$, where edge messages are accumulated according to a permutation invariant function, e.g. a mean function. The node update uses (optionally) the global state, the aggregated edge state and the current node state. Finally, a global block aggregates with permutation invariant functions the edge and node properties ($\rho^{e\rightarrow u}, \rho^{v\rightarrow u}$), and optionally uses the global state for updating the global variable state. Different parts of the full GN computation may be omitted. Several Graph Neural Network architectures can be cast as special cases of GNs by omitting certain features or by special choices of the different functions involved \cite{graphnetpaper}. In what follows, when referring to GNNs, the most general and expressive GN layer is implied except otherwise specified.

In the proposed model, entities (nodes), relations (edges) and global attributes contain both deterministic and stochastic variables.
These variables in turn, may be observable or not directly observable. Both observable and unobservable attributes may be deterministic 
or stochastic (static or evolving). In what follows, a part of the observable quantities is referred to 
as \textit{conditioning} or \textit{context}. The node, edge and global observable quantities are denoted as $\mathbf{v}^h, \mathbf{e}^h, \mathbf{u}^h$ 
where $h$ signifies that a variable corresponds to conditioning. Conditioning variables may either correspond to conditioning with known dynamic quantities 
or static quantities. Common instantiations of such conditioning are positional encoding for vertices, relative position for edges between vertices 
and time of day as a global variable. The node, edge and global variables that correspond to the rest of the states (stochastic, evolving, unobserved)
are denoted by $\mathbf{v}^d, \mathbf{e}^d, \mathbf{u}^d$.
In essence, the conditioning attributes can be used to create a \textit{conditioning graph variable}  $G_h = (\mathcal{V}_h, \mathcal{E}_h,\mathbf{u}_h) $ and a \textit{state graph variable}
$G_x = (\mathcal{V}_x, \mathcal{E}_x,\mathbf{u}_x) $.
The full graph state, is denoted by $G_d = (\V_x \cup \V_h, \E_x \cup \E_h, \mathbf{u}_x \cup \mathbf{u}_h)$ where $\cup$ denotes set union.
Since part of the node, edge and global attributes may be stochastic, a graph structured latent variable $G_z = (\mathcal{V}_z,\mathcal{E}_z, \mathbf{u}_z) $ 
is assumed. 
The graph structure may also be determined through the edge variables as in \cite{kipf2018neural}, but we restrict our model to a pre-determined 
graph structure in this work. The following model is proposed for the joint distribution of the graph structured observations
\begin{equation}
  p(\G_x; \G_h) = \int p(\G_x | \G_z; \G_h) p (\G_z; \G_h) d\G_z
\end{equation}
where $p(G_z; G_h) = p(\V_z; \V_h) p(\E_z; \E_h) p(\mathbf{u}_z; \mathbf{u}_h) $ is the distribution of the latent variables given $G_h$. 
A prior distribution conditioned on $G_h$ is assumed for the latent variable, which is further factorized along each edge and node latent separately, \ie
\begin{align}
  p(G_z; G_h) &= p^{(\V)}(\V_z;\V_h) p^{(\E)}(\E_z ; \E_h) p^{(\mathbf{u})}(\mathbf{u}_h ; \mathbf{u}_z) \\ 
  & = \prod_{i=1}^{N^v} p(\mathbf{v}^z_i ; \mathbf{v}^h_i) \cdot \prod_{k=1}^{N^e} p(\mathbf{e}_k^z; \mathbf{e}^h_k) \cdot p(\mathbf{u}^z; \mathbf{u}^h).
\end{align}
An approximate posterior (\ie \textit{recognition model}) is assumed for $ \G_z $ as $q_{\bp} (\G_z|\G_x;\G_h) $ together with a generative model for $ G_x $, $p_{\bt}(\G_x | \G_z;\G_h)$. 
In correspondence with the Variational Autoencoder (VAE) \cite{kingma2013auto}, we seek to learn the generative model parameters $\bt$ and inference model parameters $\bp$ simultaneously. 
Assuming independent identically distributed (i.i.d.) graph observations $\{ \G^{(1)}_x, \dots \G^{(i)}_x \}$, the Evidence Lower Bound (ELBO) for the marginal log-likelihood reads
\begin{align}
  \mathcal{L}(\bt, \bp; G^{(i)}_x, G^{(i)}_h) = & \mathbb{E}_{q_{\bt} (G_z | G^{(i)}_x ; G^{(i)}_h)}\big[ \log p_{\bt} (G^{(i)}_x | G_z ; G^{(i)}_h) \big] \nonumber \\ 
  &- D_{KL}(q_{\bp}(G_z | G_x^{(i)} ; G^{(i)}_h) || p_{\bt}(G_z ; G^{(i)}_h) )
\end{align}

We seek to perform fast and scalable approximate inference over the $G_z$ graph variable and at the same time take advantage of the \textit{relational structure} in the data.
A particularly convenient choice for parametrizing $G_z$ is to assume a parametric distribution over edges, nodes and globals. A GN is proposed for inferring the 
parameters. For a graph structured observation 
$G_x$, a graph structured conditioning $G_h$ and a graph structured latent $G_z$ we write
\begin{align}
  \V^{z} & \sim  q_{\bp}^{(\V)} (\G_z | \G_x; \G_h) =\, \mathcal{N}(f^{\mu_{(\V)}}_{q_{\bp}}(\G_x;\G_h),         f^{\sigma_{(\V)}^2}_{q_{\bp}}(\G_x;\G_h)) \\ 
  \E^{z} & \sim q_{\bp}^{(\E)} (\G_z | \G_x; \G_h) =\, \mathcal{N}(f^{\mu_{(\E)}}_{q_{\bp}}(\G_x;\G_h),         f^{\sigma_{(\E)}^2}_{q_{\bp}}(\G_x;\G_h)) \\ 
  \mathbf{u}_{z}   & \sim q_{\bp}^{(\mathbf{u})} (\G_z | \G_x; \G_h) =\, \mathcal{N}(f^{\mu_{(\mathbf{u})}}_{q_{\bp}}(\G_x;\G_h), f^{\sigma_{(\mathbf{u})}^2}_{q_{\bp}}(\G_x;\G_h)).
\end{align}
The functions $f^{\mu_{(\cdot)}}_\cdot$ and $f^{\sigma^2_{(\cdot)}}_\cdot$ are implemented by a GN to allow for taking into account in a general manner relational information 
while inferring over $\V_z, \E_z$ and $\mathbf{u}_z$. In practice a shared, single GN, $f_{q_{\bp}}(\cdot)$ is used. The parametrization for vertices, edges and global variables are
the corresponding states of the GN at the final message passing step. 
In a similar manner, a GN generator network, $g_{p_{\bt}}(\cdot)$, is used for $p_{\bt}$. 
Since the prior and posterior are factorized over nodes, edges and the global variable of each graph datapoint, the ELBO is split accordingly as
\begin{align}
  \mathcal{L}(\bt, \bp; G^{(i)}_x, G^{(i)}_h) = & \mathbb{E}_{q_{\bt} (G_z | G^{(i)}_x ; G^{(i)}_h)}\big[ \log p_{\bt} (G^{(i)}_x | G_z ; G^{(i)}_h) \big] \nonumber \\ 
  &- \beta_{\V} D_{KL}(q^{(\V)}_{\bp}(G_z | G_x^{(i)} ; G^{(i)}_h) || p^{(\V)}_{\bt}(G_z ; G^{(i)}_h)) \nonumber \\
  &- \beta_{\E} D_{KL}(q^{(\E)}_{\bp}(G_z | G_x^{(i)} ; G^{(i)}_h) || p^{(\E)}_{\bt}(G_z ; G^{(i)}_h)) \nonumber \\ 
  &- \beta_{\mathbf{u}} D_{KL}(q^{(\mathbf{u})}_{\bp}(G_z | G_x^{(i)} ; G^{(i)}_h) || p^{(\mathbf{u})}_{\bt}(G_z ; G^{(i)}_h)) 
\end{align}
where $\beta_{\V}, \beta_\E, \beta_{\mathbf{u}}$ can be used for controlling disentanglement as in $\beta$--VAE \cite{betavae} or the rate-distortion characteristics of the model \cite{ratedist} 
or for preventing posterior collapse and aiding training through \textit{KL-annealing} \cite{bowman2015generating,fu2019cyclical}. In a similar manner to VAEs, the approach to representing distributions over graph data with a distribution that factorizes over $\V ,\E$ and $\mathbf{u}$ allows for defining alternative evidence lower bounds for variational Bayes.
Note that the distribution does not need to be factorized along the elements of the latent vector. This allows straight-forward extensions using more flexible distributions \cite{normalizingflows}. 
A generative model based on normalizing flows that uses shift-scale transformations \cite{realnvp} has already been proposed in \cite{graphnormflow} for graph generation.
The Relational VAE (RVAE) model proposed can be extended as a hierarchical VAE \cite{sonderby2016ladder} yielding a model akin to Doubly Stochastic Variational Neural Process (DSNPV) \cite{wang2020doubly}, 
which uses global and node variables.
Finally, Neural Processes \cite{garnelo2018conditional,garnelo2018neural} (NP) and other graph encoder-decoder models \cite{kipf2018neural, tavakoli2020continuous, simonovsky2018graphvae, grover2019graphite, kipf2016variational} are closely related to the proposed model. 

\section{Related work}
\begin{figure}
  \makebox[\textwidth][c]{
    \includegraphics[width=0.9\linewidth]{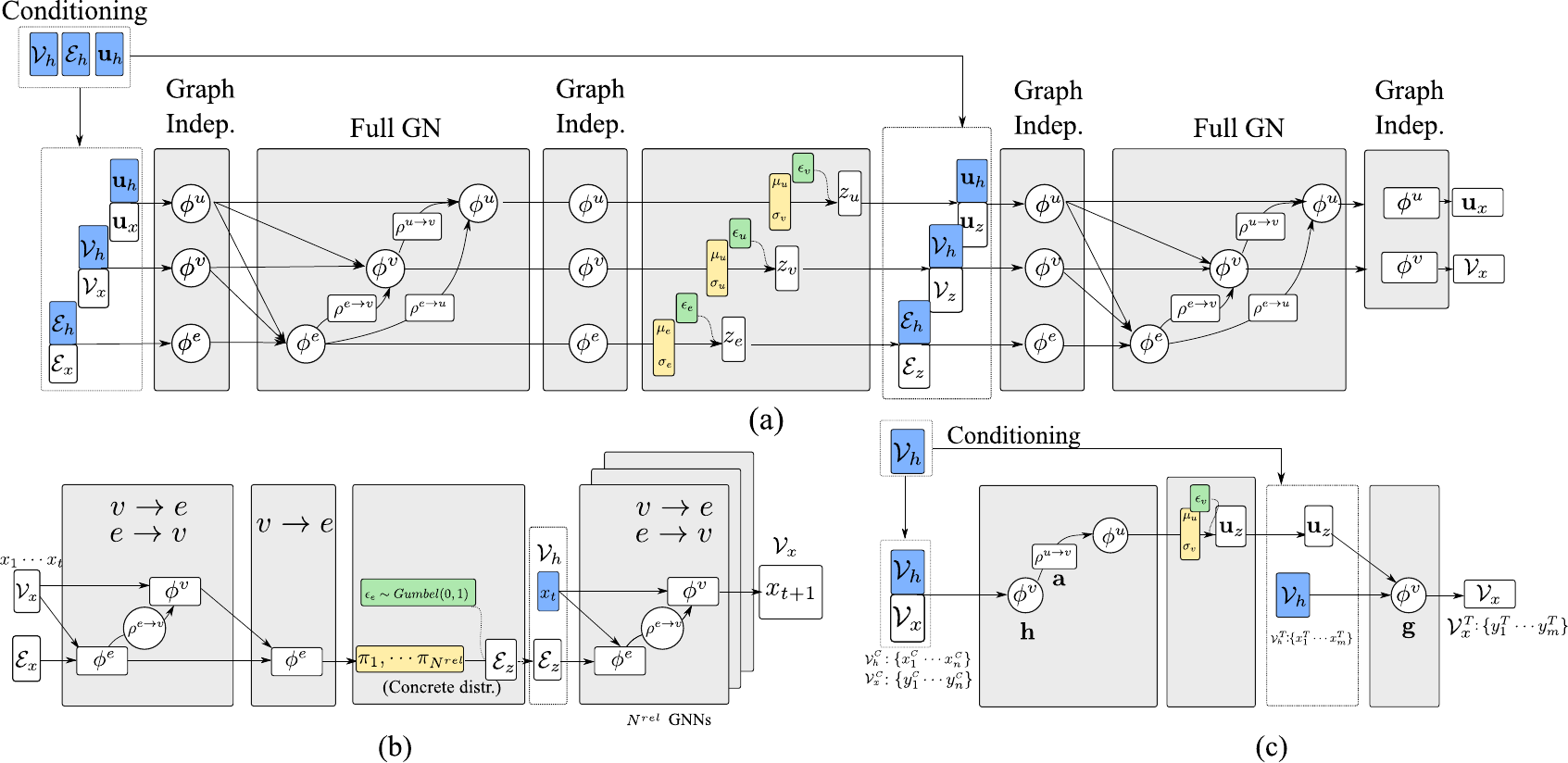}}
  \caption{\textbf{(a)} Proposed architecture with a single message passing step in the encoder and decoder \textbf{(b)} the Neural Relational Inference model of \cite{kipf2018neural}. \textbf{(c)} The Neural Process model \cite{garnelo2018neural}. For direct correspondence between the present work and \cite{kipf2018neural} and \cite{garnelo2018neural} the notations of the other works are included in the figure (e.g. $\mathbf{\rho^{v \rightarrow u} = \mathbf{a}}$ in the Neural Process model).}
  \label{fig:architecture}
\end{figure}

\paragraph{GNN Encoder-decoder models}
\label{sec:gnnreview}
In Neural Relational Inference (NRI) \cite{kipf2018neural} discrete edge latent variables are inferred from node representations and a re-parametrized $Gumbel-Softmax$ distribution is used\cite{jang2016categorical,maddison2016concrete}. 
A coarse representation of the computational graphs of NRI, NPs and the RVAE is shown in \autoref{fig:architecture}. 
In \cite{tavakoli2020continuous} graphs are modeled from global continuous
latent variables, which are subsequently used for graph generation through an adjacency matrix.
In GraphVAE \cite{simonovsky2018graphvae} the global variable together with a graph-structured conditioning variable is used 
for generation. In \textit{Graphite} \cite{grover2019graphite} a latent variable for each node is inferred from the encoder, 
while the edge variables (\ie symmetric adjacency matrix) is inferred through efficient iterated message passing.  
Similarly, the VariationalGAE\cite{kipf2016variational} uses a separate latent variable for every node and a graph convolutional encoder.
Several of the aforementioned works take advantage of recent advances in 
low-variance gradient estimates for distributions over latent variables, as in Variational Autoencoders (VAEs) via the 
reparametrization trick \cite{kingma2013auto,rezende2014stochastic}. The overlapping traits of the aforementioned are the treatment of edge, node and global variables.
In \autoref{tab:comparison} a summary of the relational modeling capabilities of various graph encoder-decoder models 
is offered. Note that the table highlights only the relevant parts to this work together and several 
important and influential design choices for graph representation learning were not touched upon.
For instance, the graph convolutional models of some of the aforementioned works offer the important 
advantage of scalability and small computation cost.

In this work, the above mentioned approaches, are generalized and unified in the proposed Relational Variational Autoencoder 
(RVAE) model. Note that it is not difficult to yield explicit graph connectivity in RVAE as in NRI \cite{kipf2018neural} since the type and existence of a connection can be 
seen as a categorical variable. See also \autoref{fig:architecture} (b), where a sketch of NRI is offered. Inferring graph connectivity or generating graphs, however, falls out of the scope of this work.
In RVAE the focus is generative modeling of graph structured data with an apriori known connectivity, with attributed nodes and edges, which optionally may include a global attribute that influences both entities and relations. 

\begin{table}[h!]

\footnotesize
{

  \caption{Features of different related Bayesian graph network encoder-decoder models (see also \autoref{fig:architecture}). 
  For the NP models that contain a latent variable, it is straightforward to combine a deterministic global encoder for the context inputs at test time \cite{dubois2020npf}. 
  The attributes with subscript $z$ denote that the model performs optimization using an ELBO objective. The attributes with superscript $h$ denote whether the models may facilitate deterministic conditioning for the corresponding graph attribute at test time. }
  \label{tab:comparison}
  \centering
  \begin{tabular}{r ccc ccc l}
    \toprule
     &\multicolumn{3}{l}{Latent} & \multicolumn{3}{l}{Conditioning}   & \\
    \cmidrule(r){1-8}
    Name                                   & $\mathcal{V}_z$ & $\mathcal{E}_z$ &$\mathbf{u}_z$ &$\mathcal{V}_h$& $\mathcal{E}_h$ &$\mathbf{u}_h$ & Architecture notes \\
    \midrule
    CNP \cite{garnelo2018conditional}      & \xmark &  \xmark & \xmark &\cmark & \xmark &\cmark   & DeepSet encoder, GN node block \\
    AttCNP \cite{kim2019attentive}         & \xmark &  \xmark & \xmark &\cmark & \cmark &(\cmark) & Attention encoder/decoder\\
                                           &        &         &        &       &        & &          Decoder edge cond. through cross-attention \\
    ConvCNP \cite{gordon2019convolutional} & \xmark &  \xmark & \xmark &\cmark & \cmark &(\cmark) & SetConv encoder     \\
    NP \cite{garnelo2018neural}            & \xmark &  \xmark & \cmark &\cmark & \xmark &(\cmark) & DeepSet encoder \\  
    GraphVAE \cite{simonovsky2018graphvae} & \xmark &  \xmark & \cmark &\cmark & \cmark &\cmark   & Graph conv. encoder \\  
    VariationalGAE \cite{kipf2016variational} & \cmark&\xmark & \xmark &\xmark & \xmark &\xmark   & Graph convolutions  \\
    Graphite \cite{grover2019graphite}     & \cmark &  \xmark & \xmark &\cmark & \xmark &\xmark   & Iterative decoder   \\
    NRI \cite{kipf2018neural}              & \xmark & \cmark & \xmark &\cmark & \xmark &\xmark   & MP encoder/decoder\\  
    MPNP \cite{day2020message}             & \xmark & \xmark & \cmark &\cmark & \cmark &(\cmark) & MP encoder/decoder  \\ 
    DSVNP \cite{wang2020doubly}            & \cmark & \xmark  & \cmark &\cmark & \xmark & (\cmark) & $\mathcal{V}^z\sim p(\mathcal{V}^z|\mathbf{u}^z, \mathcal{V}^*, \mathcal{V}^{h*})$\\
    RVAE (this work)                       & \cmark & \cmark & \cmark &\cmark & \cmark &\cmark   & MP encoder/decoder  \\ 
    \bottomrule
  \end{tabular}
}
\end{table}

\paragraph{Neural processes}
\label{sec:nps}
In Neural Processes (NP)\cite{garnelo2018conditional, garnelo2018neural}, we consider a set of mappings $F: \mathcal{X} \rightarrow \mathcal{Y}$ where $ \mathcal{X} : \{ x_i\},\, x_i \in \mathbb{R}^{N_x}, \mathcal{Y} : \{y_i \},\, y_i \in \mathbb{R}^{N_y}$.
A particular draw of a function $ f \sim F $, is modeled as $f(x_i) = g_\theta(x_i,z) $ where $z \sim p(z)$ is a high dimensional random vector 
(e.g. a standard normal) and $g_\theta$ is a neural network and $\theta$ denotes the parameters of $g$.
Given a set of $n_m$ input-output observations $\mathcal{D} :\{ (x_{1:n_m}, y_{1:n_m})_{f_m}\}$
from $m$ different realizations of $f$ (potentially different in number), we want to learn
a distribution over $z \sim p(z|\mathcal{D})$.
Under the NP approximation, assuming observation noise $y_i \sim \mathcal{N}(g_{\theta}(x_i,z), \sigma^2)$,
the distribution of $y$ is defined as
\begin{equation}
  p(z,y_{1:n}|x_{1:n}) = p(z) \prod_{j=1}^n \mathcal{N}(y_i | g(x_i, z), \sigma^2).
\end{equation}
In practice, the input-output observation cases $\mathcal{D}$, are split as $\mathcal{D}^{C \cup T} = \mathcal{D}^C \cup \mathcal{D}^T$, where $C$ denotes a set of points 
with observations in $\mathcal{X}$ and $\mathcal{Y}$ and $T$ denotes a set of points where we only observe $\mathcal{X}$ (\ie the inputs).
This can be cast as a conditional generative model for $p(y_T | x_T,x_C, y_C) = p_\theta(y_T|x_T,z)p(z|x_C,y_C)$, where the conditioning is the fully observed context pairs.
The ELBO used for optimization is
\begin{align}
  \label{eq:nploss}
  \log p(y_T|x_T,x_C,y_C) \geq & \mathbb{E}_{q_{\phi}(z | \mathcal{D}^{C \cup T})}\Big[ \sum_{i \in T} \log p_\theta(y_i|z, x_i) + \log \frac{q(z|\mathcal{D}^C)}{q(z|\mathcal{D}^{C \cup T})} \Big] \nonumber \\
                          %& \mathbb{E}_{q_{\phi}(z | \mathcal{D}^{C \cup T})}\Big[ \sum_{i \in T} \log p_\theta (y_i|z, x_i) \Big] - \mathbb{E}_{q_{\phi}(z | \mathcal{D}^{C \cup T})}\Big[ \log \frac{q(z|\mathcal{D}^{C\cup T})}{q(z|\mathcal{D}^{C})} \Big]  \nonumber  \\
                          & \mathbb{E}_{q_{\phi}(z | \mathcal{D}^{C \cup T})}\Big[ \sum_{i \in T} \log p_\theta(y_i|z, x_i) \Big] - D_{KL}(q_{\phi}(z | \mathcal{D}^{C \cup T}) || q_{\phi}(z | \mathcal{D}^C)) 
\end{align}
Note that the above variational objective has an intuitive interpretation, as a reconstruction loss (first part) 
and a Kullback-Leibler divergence between the approximate posterior distributions predicted when using both $C \cup T$ and when 
using only $C$ (the context set).
In \cite{ivanov2018variational} a similar loss function was proposed with the motivation 
of training VAEs that can be used with arbitrary conditioning masks. By considering the set of observations as nodes in 
a disconnected graph, (\ie $\mathcal{V} : \{ \mathbf{v}_i | (x_i, y_i) \}_{i = 1:N^v}$) and training while masking the context 
output nodes $y_C$, the same objective is retrieved. 
Therefore, following the nomenclature of \cite{graphnetpaper}, we can instantiate a NP from the proposed model, by using arbitrary conditioning as described in \cite{ivanov2018variational}, 
a DeepSet \cite{zaheer2017deep} as an encoder and only a node-block as a decoder as shown in \autoref{fig:architecture}.

The NP framework has been extended to take advantage of special inductive biases, such as the \textit{relation} of observation and target nodes in Attentive Conditional Neural Processes (AttNP) \cite{kim2019attentive} or the translation equivariance in Convolutional Conditional Neural Processes (ConvCNP) \cite{gordon2019convolutional}.
More recently, relational inductive biases were employed in Message Passing Neural Processes (MPNP) \cite{day2020message}. 
The aforementioned models, feature a global latent variable $\mathbf{u}_z$ which is inferred from the context points and parametrizes the distribution over functions.
With the exception of MPMP, the aforementioned works target non-relational data. Nevertheless, MPMPs does not directly implement edge-bound uncertainty or edge-level conditioning, which is the most pronounced difference to RVAE. Similar to this work, in DSNPV \cite{wang2020doubly} a NP that allows for both node and global latent variables was proposed, which in addition, employs a hierarchical VAE \cite{sonderby2016ladder}. The motivation of DSNPV is to include node-context information, which in the conditional RVAE is also supported by design through $\mathcal{V}_h$. RVAE attempts to merge the complementary strengths of the aforementioned models in representation of uncertainty, with a focus towards modeling graph structured continuous data. Finally, in contrast to Functional Neural Processes \cite{fnp} we do not deal with inferring a graph of dependencies among latent variables, yet hierarchical RVAE adaptations may also manage such tasks.
% At the core of the approach, is treating graph-valued (\ie  \textit{relational}) data as \textit{first-class citizens} and provide algorithms with global, node and edge latent variables and conditioning. 

%\paragraph{Generative models and graph ML}
%In contrast to vanilla VAEs, which are latent variable models defined on data with no relational structure, RVAEs operate in relational (i.e. graph) data.
%Correspondingly, in contrast to NPs which learn a distribution over functions of Euclidean data, a Conditional RVAE with arbitrary conditioning as described in \autoref{sec:rvaeac} 
%allows for learning a distribution over graph neural network functions that operate on graph data (seel also \autoref{sec:nps}).

%%%%%%%%%%%%%%%%%%%%%%%%%%%%%%%%%%%%%%%%%%%%%%%

\paragraph{Graph Gaussian processes}
Sharing the motivation of this work, \ie taking advantage of relational information and learning joint distributions of graph structured data,
in \cite{sindhwani2007relational} GPs were defined over graphs with undirected binary (positive or negative) edges and applied to semi-supervised learning problems.
In \cite{ng2018bayesian} the authors applied GPs trained with variational approximations for semi-supervised learning on graphs that contain non-attributed edges.
In \cite{deepgraphgp} GP-based approaches are fused with deep learning for learning graph (e.g. network) structured signals. 

%in the literature include \cite{deepgraphgp} where 
%a Gaussian process approach was proposed for learning signals over  graphs. 
%The difference between this work and CNPs/NPs is that the proposed model allows for learning 

%In \cite{kipf2018neural} trajectories of interacting entities, $\{X_1 \cdots, X_t \}$ with $X_i$ denoting observed entity states, 
%are processed by GNN to yield, after message passing, a categorical distribution over possible edge types.
%The decoder consists of a different message passing neural network for every different type of edge, including a \textit{no-edge} category
%which denotes the absence of edge between two entities. The last observation of the trajectories, $X_t$ is used to construct the node features of the decoder, which in turn is used to predict
%the next step $X_{t+1}$. 

%Consider a set of observations from a function $F: \mathcal{X} \rightarrow \mathcal{Y}$ with the set of input points 
%$x:\{x^C \cup x^T\} \subset \mathcal{X} $ and $ y:\{y^C \cup y^T \} \subset \mathcal{Y}$ the set of output points, and the superscripts $\square^C$ and $\square^T$ denote the \textit{context} and \textit{target} points correspondingly. 
%In what follows, the RVAE model is adapted to be trainable with special conditioning, so as to allow for its usage as a Neural Process, 
%(i.e. a distribution over functions learnable directly from observations) by applying conditioning on a set of context observations $\{x^C,y^T\}$ and a set of target $x^T$ inputs.
%A set of context observations $s^C \{ x^C,y^C\}$ and a set of target observations $s^T\{ x^T, y^T \}$ are defined, where $y = f(x)$ and $f \sim p(f;\mathcal{C})$

%The sets of context 
%In \cite{ivanov2018variational} a method for defining a Variational Autoencoder with Arbitrary Conditioning (VAEAC) is proposed.
%The sets of 
%Node features are defined 
%The RVAE is adapted to support 

\section{Results}
\subsection{Wind farm operational data}
A real-world industrial application, where relational structure is inherent in the observed data, 
is found in modeling of operational data of wind turbines positioned in a farm. The wind turbines (nodes) feature static 
variables, such as their power production characteristics and their position, as well as dynamic variables such 
as their current operational state. The actual operational state of a turbine is only known up to a
certain precision from historical data, (\ie Supervisory Control and Data Acquisition (SCADA) data),
which is usually limited to 10 minute statistics. Due to the stochasticity of the wind excitation, 
compounded by incomplete information due to coarse measurements, there is \textit{uncertainty} 
associated with the actual operational state of a wind turbine. Wind turbines arranged in a wind 
farm interact through the so-called \textit{wakes}, which are travelling vortices that affect the 
power production and vibrations of downstream turbines. The magnitude of wake effects is related 
to large scale turbulence (which is a global dynamic variable), to wind orientation (which is a global 
dynamic variable), to upwind turbine nacelle orientation (which is a node dynamic variable), the relative 
position between the two turbines (an edge static variable), the rotor diameter and the distance between 
the two turbines. The interaction is one-way directional but can change directionality depending on the 
wind orientation. The effect of wakes is stochastic due to turbulence. For robust wind power prediction, 
monitoring, control, and maintenance planning, we want to infer the distribution of operational characteristics 
of a wind farm conditioned on turbine characteristics and farm layout.
Of crucial importance is the
inclusion of stochastic variables in the interactions (\ie edges) of the considered graph. 
Static graph edges, used as part of the graph conditioning, are constructed by considering the spatial proximity and relative position of pairs of turbines. 
The goal is to generalize directly to unseen farm configurations while learning directly on real condition monitoring data 
(zero-shot generalization) but at the same time to yield uncertainty estimates. 

\paragraph{Graph machine learning in wind farm modeling} In~\cite{park2019physics} a GNN was trained on simulated data for wind power prediction. Recently, in~\cite{bleeg2020graph} GNNs were applied as a surrogate model to more accurate fluid dynamic simulations. With the architectural advancements proposed in this work, we extend the wind farm relational modeling literature by providing a solution for representation of uncertainty in wind turbine interactions. Moreover, we empirically show in real wind farm data that significant accuracy improvements are possible through the incorporation of the proposed relational modeling and variational Bayes approach.

\begin{figure}[h!]
  \includegraphics[width=\textwidth]{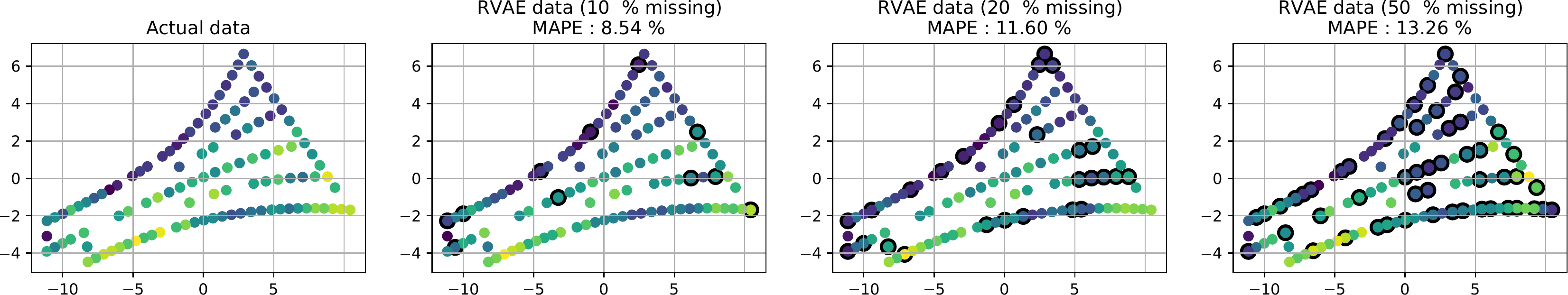}
  \caption{Imputation qualitative results for wind speed. The imputed points are marked with a dark circle on the background. The mean absolute percentage error is reported, which is computed as $1/N^T\sum_{i=1}^{N^T} (|\vi^T - \hat{\vi}^T|) / |\vi^T| $ where $\vi^T$ is the actual value of node $i$ , $N^T$ the number of target turbines and $\hat{\vi}$ is the CRVAE prediction.}
  \label{fig:imputation}
\end{figure}

\subsection{Real wind farm SCADA dataset}
\label{sec:realscada}
Conditional RVAE models (CRVAE) were trained with with a 80/20 train/test split on a dataset that includes 6 months of 10-minute average SCADA data readings. Since the goal is to compare the fitting capability of the models and not model selection, no validation set was employed. Early stopping with patience of 2500 steps was used (test set evaluation every 500 steps). The larger RVAE models that also yield the best performance had not converged at the 10th epoch.
The 20\% of turbine data are randomly masked during training. A batch size of 16 was used for all models. In order to make a fair comparison no regularization or KL-annealing was used. A small learning rate of $5\cdot10^{-5}$ and the Adam optimizer \cite{kingma2014adam} with default
parameters was used for all the runs.
The final ELBOs for all models are shown in \autoref{tab:elboanholt}. A $mean$ aggregation function and composite aggregation function consisting of a concatenation of $mean$,$max$ and $min$ aggregators were used. Due to the concatenation operation, the composite aggregators result in slightly larger networks. Aligned with recent results on GN performance when using composite aggregation functions \cite{pna} we find that networks with the $mean-max-min$ aggregator indeed yield better performance. The motivation, however, for using composite aggregators, is also due to the physics of the problem. By using such aggregators it is easier to discriminate the un-waked part of the farm and the waked turbines. More concretely, turbines at the upstream boundary of the farm have larger power production and this directional effect can easily be masked using the mean aggregation. The CRVAE models are compared to a two-layer MLP-based CVAE trained with the arbitrary conditioning objective \cite{ivanov2018variational} of varying sizes, with the largest CVAE model number of parameters corresponding to the number of parameters of the best performing RCVAE. The largest CVAE model was the worst-performing of the evaluated CVAE models.

The CVAE model with the smallest size has slightly better performance compared to the RVAE model that performs no message passing on the encoder part, and therefore ignores relational inductive biases when inferring $G_z$. 
All but one of the CRVAE models strongly outperform the CVAE models by a large margin which is attributed to the effective use of relational inductive biases. To further support this claim, in the supplemental material (\autoref{sensitivities}) gradient sensitivities are plotted and it is observed that the imputation results for masked turbines depend on upstream turbines. Qualitative imputation results are shown in \autoref{fig:imputation}.

\begin{table}[h!]
  \centering
\footnotesize{
  \caption{Test set ELBO on Anholt SCADA dataset after 10 epochs. Numbers in parentheses are the standard deviations of the ELBO estimates in the test set (higher is better).
  The same node, edge and global latent sizes were used ($N^{G_{z}}$). ``(comp.)'' stands for the composite mean-max-min aggregator. All MLPs are 3 layer ReLU MLPs. The $\cdot^*$ superscript denotes results that were not derived from early stopping.}

  \label{tab:elboanholt}
  \centering
  \begin{tabular}{r cc c c c c c}
    \toprule
     %&\multicolumn{3}{l}{Latent} & \multicolumn{3}{l}{Conditioning}   & \\
           &           &             & \multicolumn{2}{l}{MP Steps}  & & \\
               \cmidrule(r){4-5}
     Model & mlp units & $  N^{G_{z}}$ size  & enc. & dec. & agg. & \# params & ELBO \\
    \midrule
      CRVAE   & 64 & 32 & 0 & 1 & mean &      184,717  & $1.96, (0.30)$ \\ 
              & 64 & 32 & 1 & 1 & mean &       341,517 & $6.99 (0.29)$\\
              & 64 & 32 & 2 & 2 & mean &       498,317 & $7.48 (0.61)^*$\\ 
              & 64 & 32 & 2 & 2 & (comp.)&     522,893 & $\mathbf{8.11 (0.48)^*}$\\ % (-7.2, (0.46))
              & 64 & 32 & 3 & 3 & (comp.)&     679,693 & $7.70 (0.53)^* $ \\ % (-6.68, 0.42)
    \midrule
    CVAE      & 128 & 64  & -- & -- & -- &  77,194     & $ 2.12 (0.10) $\\ %(-0.29953775, 0.114754386),
              & 256 & 64 & -- & -- & -- &  252,554     & $ 1.17 (0.16) $ \\
              & 384 & 96 & -- & -- & -- &  563,146     & $ 1.23 (0.09)$ \\ 
    \bottomrule
  \end{tabular}
}
\end{table}

{
\begin{wraptable}{R}{5.5cm}
%\begin{table}
  \footnotesize{
    \caption{Effect of edge latent variables. Results based on 3 runs for each case.}
    \label{tab:edgeablation}
    \centering
    \begin{tabular}{r c c}
      \toprule
      Case &        $\log{p(\hat{\V}_x | G_z; G_h)} $ & Range  \\ 
      \midrule
      $\beta_{\E} = 1.$ & $\mathbf{4.16}$ & $\pm 0.43$ \\
      $\beta_{\E} = 0.$ & $1.80$ & $\pm 1.21$ \\ 
      \bottomrule
  \end{tabular}
%\end{table}
}
\end{wraptable}
\paragraph{Effect of inferring edge latents $\E_z$} The introduction of continuous edge-related latent variables is overlooked in a large part of the literature. Wake effect modeling is an application that may benefit from edge latent variables. We test the effect of edge latent variables by setting $\beta_{\E} = 0$ while still using $G_h$. The results of this experiment are shown in \autoref{tab:edgeablation}. The inclusion of the KL term with respect to edge latent variables seems to improve the reconstruction error achieved by the model.

\subsection{Wind farm simulation dataset}

The steady-state wind farm wake simulator FLORIS \cite{floris} was used. A dataset of wake effect simulations and preprocessing tools for demonstrating the wind farm modeling approach adopted herein is released as part of this work. In what follows we test the generalization capabilities of a trained RVAE to novel geometric configurations.
A single farm configuration is used for training and another one is used for testing. Both farms are simulated with random wind characteristics such as direction and average wind speed. An example output from the simulation can be found in the supplemental material.
The train and test farm configurations can also be found in the supplemental material. 
\paragraph{Qualitative results} The RVAE model is able to capture the orientation-dependent wake 
deficit for each turbine separately on the test wind farm as shown in \autoref{fig:traintestfarm_deficits}. Furthermore, we use a single turbine as a probe and position it on a regular grid while keeping a turbine on a fixed position (0,0).
By inspecting the wind speed predicted at the probe turbine, we can map the wake deficit in 2D behind the source turbine. 
This is shown in \autoref{fig:learnedwindwake}. The spatial dependence of the wake deficit is also shown as computed from FLORIS and the error in RVAE estimation.
For distances larger than $200m$ the wind deficit is accurately predicted. 
Note that this result is from a model trained on operational data from a \textit{single} simulated windfarm.
When the turbines are very close ($<200m$) the wakes are not predicted correctly, but this is an expected effect since the RVAE never encounters turbines at these distances.
Wake effects estimated with the RVAE are slightly lower than those derived from the simulation as shown in \autoref{fig:traintestfarm_deficits}. However, the RVAE seems to capture the intricate wind orientation-dependent effects which depend on the farm layout.

%\begin{wrapfigure}{r}{5.5cm}
\begin{figure}
  \centering
  \includegraphics[width=1.\textwidth]{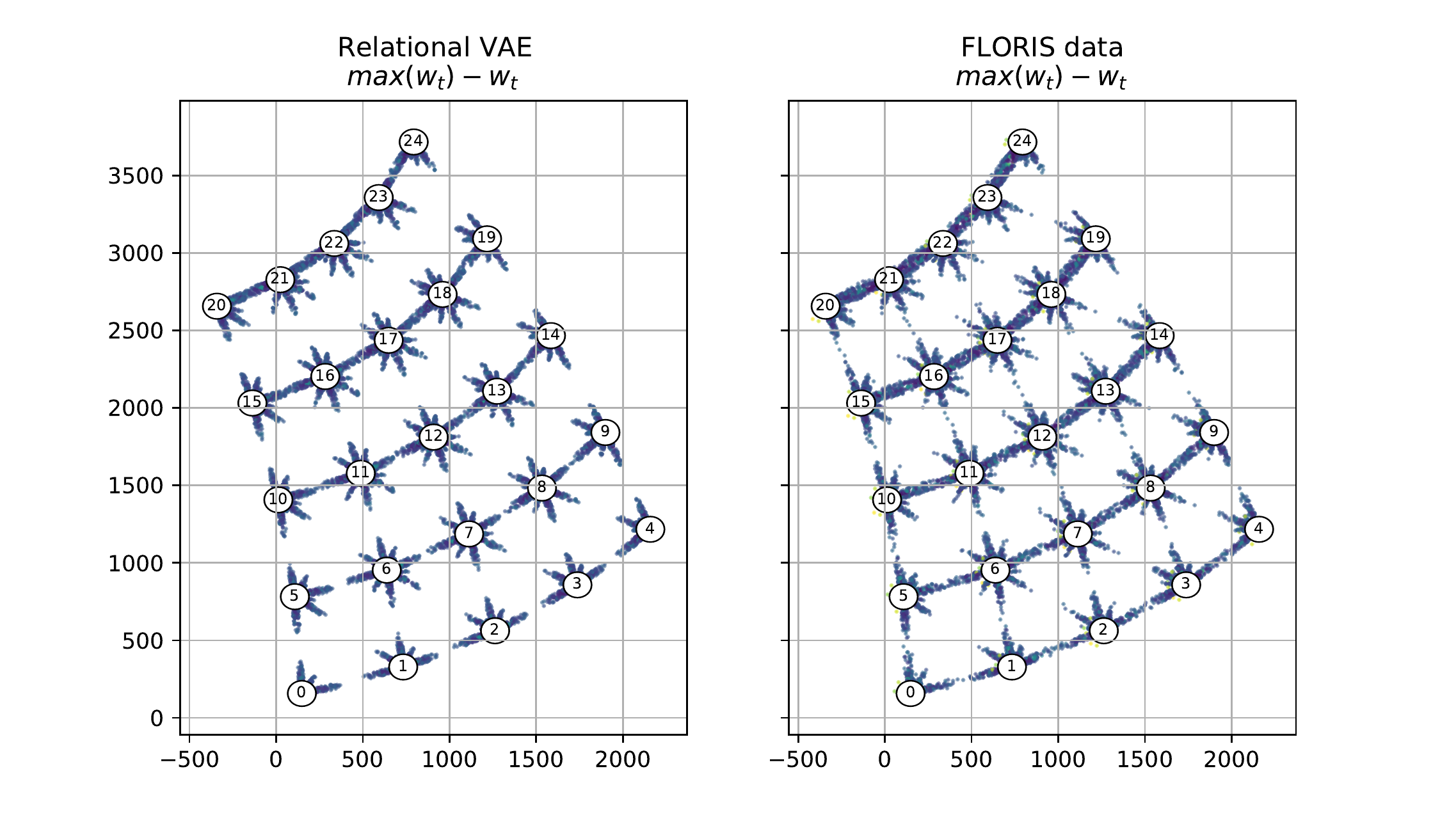}
  \caption{Wind deficits on the simulated test farm and estimates from the trained RVAE. Each point associated with a turbine is plotted in a 2D polar coordinate system centered on the turbine. Each point is plotted towards the orientation of the \textit{incoming} wind. The distance from the origin is proportional to the wake deficit, estimated as $max(v) - v$ where $v$ is mean power and mean wind. }
  \label{fig:traintestfarm_deficits}
\end{figure}
%\end{wrapfigure}

\begin{figure}
  \centering
  \includegraphics[width=1.0\textwidth]{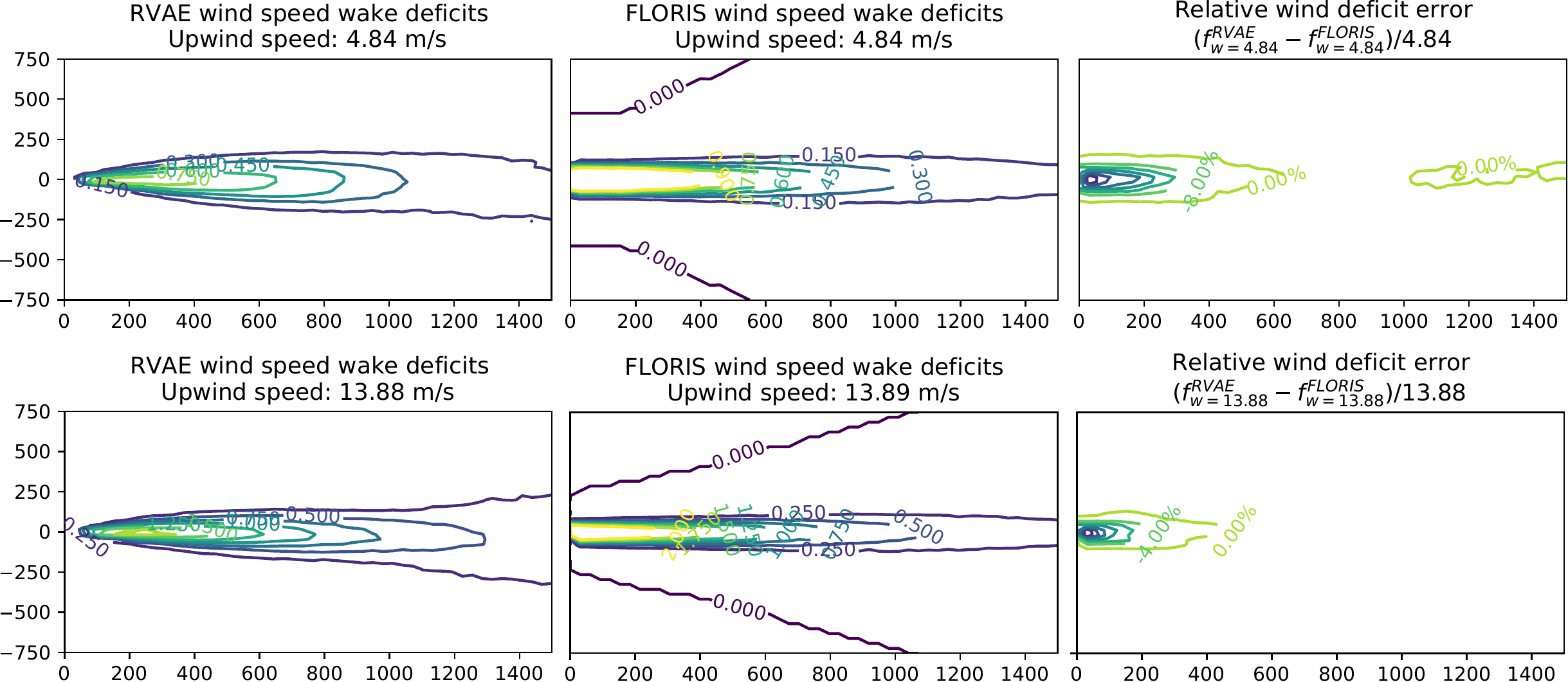}
  \caption{Learned spatial distribution of wake related wind speed deficit, evaluated as $w_{(0,0)} - w_{(x,y)} $ where $w_{(0,0)}$ is the the wind speed at the up-wind turbine and $(x,y)$ denotes the wind speed for a \textit{probe} turbine positioned at $w_{(x,y)}$.}
  \label{fig:learnedwindwake}
\end{figure}

\subsection{1D regression}
\label{sec:rvaeac}
In order to further demonstrate the versatility of the RVAE in modeling structured data, and in order to make the connection 
to NPs \cite{garnelo2018neural} clearer, in what follows an RVAE adapted for node data imputation is presented \cite{ivanov2018variational}.
The dataset consists of sets of points sampled from a zero-mean 1D Gaussian process with a squared exponential kernel. The 
pairs of input points $\{ (x_{1:n_m}, y_{1:n_m})_m \} $ are used as node features to construct a set of context and target graphs, where $m$ corresponds to 
different GP realizations. Each graph contains a set of edges $\mathbf{e}_i$ which encode the relative position of the observation points. 
The edge features between observations at points $x_i, x_j$ are defined as $f(x_i, x_j) = e^{-c\cdot |x_i-x_j|^2}$ where $c$ is a function of the cutoff distance for edge creation. 
Note that the construction of such edge features endows the model with translation equivariance.
In contrast to MPNPs, \cite{day2020message}, the edges are the same in the context and target graphs. 
The loss function used is the same as in \autoref{eq:nploss}. Both $p_\phi$ and $q_\theta$ are implemented as 
GNs. The outputs of the GNs parameterize a Gaussian, \ie
\begin{equation}
  q_\phi(z | \mathcal{D}) = \mathcal{N}(\mu_\phi(\mathcal{D}), \sigma_\phi^2(\mathcal{D})) , \quad
  p_\theta(y | \mathcal{D}) = \mathcal{N}(\mu_\theta(\mathcal{D}), \sigma_\theta^2(\mathcal{D})).
\end{equation}
The $y$ values of $\mathcal{D}^T$ are replaced with $0$ when fed through the encoder and an additional binary feature $b$ for the node, which denotes masking, is appended to the node tuple. 
The $b$ feature is zero for the unmasked nodes and $1$ for the masked nodes. The masked input is denoted by $\mathcal{D}^{T\setminus b}$. 
The union of the masked target input with the context dataset is denoted by $ D^{T\setminus b \cup C}$.
Instead of using two different functions for the prior of $p(z|\mathcal{D}^{T\setminus b \cup C})$ as 
in \cite{ivanov2018variational}, and posterior network $q(z|\mathcal{D}^{T \cup C})$ and in order to keep the conditional RVAE model closer to the NP formulation, 
the approximate posterior (\ie the encoder of the RVAE) is used also for the learned prior. 
The decoder $p_\theta$ receives as node conditioning (and optionally edge conditioning) the $x_T$ values and the global
latent variable $\mathbf{u}^z$. Each realization of $\mathbf{u}^z$ corresponds to a different context set which in turn 
corresponds to a different sampled $GP$. More information about the training setup can be found in the appendix.

%All experiments were performed using the Adam optimizer \cite{kingma2014adam} with a learning rate of $10^{-4}$ and default parameters. All RVAE models were trained for $4\cdot10^4$ steps and NP models up to $5 \cdot 10^4$ steps with batches of size 16. Each batch contains a random number of context and target points which varies between 3 and 50.
%All GN functions involved are feed-forward 3-layer Multi-Layer Perceptron (MLPs) with rectified linear unit non-linearities and no activation in the last layer.
%The GN block used is encode-process-decode architecture as in \cite{graphnetpaper} with residual connections. The encoder and decoder contain layers which do not perform message passing but only cast the inputs to a predefined dimension (Graph Independent layers).

The NP is implemented by defining a DeepSet encoder, a global latent variable of the same size as the NP MLP.
The same latent variable size for nodes and edges was used for each experiment, which is the same as the core size. All aggregation functions are \textit{mean} aggregations. 
Experiments were performed with different number of message passing steps, and inclusion of either the relative observation position as 
an edge feature or the absolute node position $x_i$ for each feature. The models are tested in un-seen GP realizations and the negative 
log-likelihood of predictions are reported in \autoref{tab:nllresults}. The RVAE models compute edge, node and global variables.
The test datasets contain points with $x \in [0,1]$ and $x \in [1,2]$ ranges in order to test the generalization capability of the proposed model in translation.
Since the edge-blocks only ever receive translation equivariant inputs from the dataset, the RVAE models generalize well in the $x \in [1,2]$ range. This is presented only as an example of how special equivariant inductive biases may be implemented in RVAE. It is observed that the full RVAE model does not perform well when only the node features are available. As with NPs, it was empirically found that models yield better results with more training. 

  \begin{table}[h!]
  \centering
\footnotesize{
  \caption{Test set log likelihoods on 1D GP regression with Conditional RVAE. The results are based on a set of 5000 unseen GP samples, each with 50 context and 50 target points. The models were trained only on points with x in the $[0,1]$ range. Values in parentheses are standard deviations of the mini-batches. RVAE denotes a model where all latent variables are used (edge node and global).}
  \label{tab:nllresults}
  \begin{tabular}{r ccccc}
    \toprule
     %&\multicolumn{3}{l}{Latent} & \multicolumn{3}{l}{Conditioning}   & \\
     Model & size & \multicolumn{2}{c}{Only cond. on nodes }& \multicolumn{2}{l}{ Cond. on edges and nodes  }\\
      &      \small{(mlp/$z$/MP Steps)} & \multicolumn{2}{c}{$G_h = (\mathcal{V}_h, \cdot, \cdot)$} & \multicolumn{2}{c}{ $G_h=(\mathcal{V}_h, \mathcal{E}_h , \cdot )$ }  \\
    \multicolumn{2}{l}{} &  $ x \in [0,1]$  & $ x\in [1,2]$         &  $x \in [0,1]$ &  $x \in [1,2]$\\
    \midrule
     CRVAE     & 64/64/0    &   $-17.94 (3.11) $   &  $ -24.59 (4.10) $    &        $ 0.33 (0.04) $   &           $-0.21 (0.13)$    \\
              & 64/64/1    &   $-12.55 (2.51) $   &   $ -9.79 (2.51) $    &  $0.36 (0.07) $    &   $ 0.08 (0.06) $    \\  
              & 64/64/2    &   --            &      --              &     $ \mathbf{0.98 (0.09)} $      &   $ \mathbf{0.67 (0.08)} $ \\
    \midrule
    NP        & 64/64/NA   &  $-1.34 (0.07)$   & $ -11.13 (3.08) $  & NA & NA \\
              & 128/128/NA &  $-1.08 (0.11) $ & $-31.74 (14.08)$ & NA & NA \\
    \bottomrule
  \end{tabular}
}
\end{table}

\section*{Conclusions and broader impact}
This work introduces an attributed graph approach to the probabilistic modeling of relations within entities and their properties.
The approach is verified and validated on wake effect simulations and actual data from wind turbines placed within a wind farm; a characteristic example that may be modeled as a graph. We also find some connections to the NP literature which we demonstrate by adapting the proposed method to perform a typical NP benchmark which is 1D regression for GP data.

We introduce a method for data-driven wake effect modeling for wind farms that accounts for uncertainty. The proposed method fuses physical intuition, flexible function approximation through GNs, and variational Bayes through re-parametrized gradients.
Better and more computationally efficient wake effect modeling can lead to improvements in terms of accuracy and computational efficiency in analysis for wind farm siting~\cite{lundquist2019costs} farm layout optimization~\cite{kirchner2019multi}, wind farm control optimization~\cite{howland2019wind} and ultimately power production improvements, as well as more robust to uncertainties maintenance planning. Ultimately, the aforementioned lead to wind energy being a more attractive clean energy solution.

Graph data are naturally used to model social, transportation and communication networks. Possible negative implications of any graph ML work relate to possible malicious uses of analysis in such networks, such as de-anonymization in social networks~\cite{ji2016graph}, and vulnerability exploitation on transportation networks. 

\begin{ack}
% Use unnumbered first level headings for the acknowledgments. All acknowledgments
% More information about this disclosure can be found at: \url{https://neurips.cc/Conferences/2021/PaperInformation/FundingDisclosure}.
% 
% Do {\bf not} include this section in the anonymized submission, only in the final paper. You can use the \texttt{ack} environment provided in the style file to autmoatically hide this section in the anonymized submission.
The authors would like to gratefully acknowledge the support of the European Research Council via the ERC Starting Grant WINDMIL (ERC-2015-
StG \#679843) on the topic of Smart Monitoring, Inspection and Life-Cycle Assessment of Wind Turbines. 
\end{ack}

%\section*{References}

% References follow the acknowledgments. Use unnumbered first-level heading for
% the references. Any choice of citation style is acceptable as long as you are
% consistent. It is permissible to reduce the font size to \verb+small+ (9 point)
% when listing the references.
% Note that the Reference section does not count towards the page limit.
\medskip
\small{
  \bibliographystyle{abbrv} 
  \bibliography{bibliography}  %%% Remove comment to use the external .bib file (using bibtex).
}

\appendix
\section{Appendix}

%Optionally include extra information (complete proofs, additional experiments and plots) in the appendix.
%This section will often be part of the supplemental material.

\subsection{RVAE on wind farm monitoring data}
The steady-state wind farm wake simulator FLORIS was used \cite{floris} for the simulated dataset. 
An example output of a simulation from FLORIS is shown in \autoref{fig:florisout}.
\begin{figure}[h]
  \centering
  \includegraphics[width=0.75\textwidth]{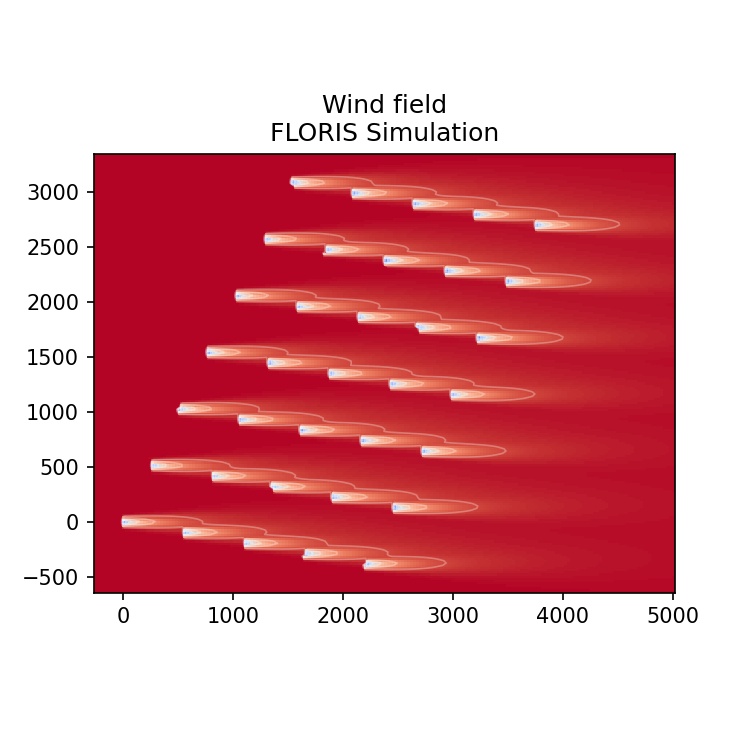}
  \caption{A representative simulated wind field from FLORIS. Lighter colors represent lower wind speeds.}
  \label{fig:florisout}
\end{figure}

\paragraph{Additional details on farm graph construction}
A graph is constructed where turbines are the vertices and directed edges are created between the turbines by truncating an all-to-all graph to a cut-off distance of $100\cdot d$ where $d$ is the turbine rotor diameter. The 5MW NREL prototype turbine was used for the simulations \cite{nrel5mw} which has a diameter $d=126m$.
In order to test the generalization capabilities of the model, and most importantly the ability to implicitly learn how to use the relative position
of the turbines and the nacelle wind orientation and power production of up-wind turbines, the model is tested on a different farm configuration than the one it is trained on.
It is noted that the wake simulator used is not stochastic. 
The yaw directions of the turbines were randomly perturbed with $\mathcal{N}(0.,5.^\circ)$ around the global wind orientation in order to introduce some stochasticity. 
The training dataset consists of $4462$ farm SCADA readings for the whole farm.
Both farms are simulated with the same randomly sampled mean wind and wind orientation global conditions ($u_h$). Due to the different arrangement of the turbines in the two farms and their wake interactions the wind speed and power production of the two farms is very different.
The farm configurations are shown in \autoref{fig:traintestfarm}.
\begin{figure}[h!]
  \centering
  %\fbox{\rule[-.5cm]{4cm}{4cm} \rule[-.5cm]{4cm}{0cm}}
  \includegraphics[width=0.8\textwidth]{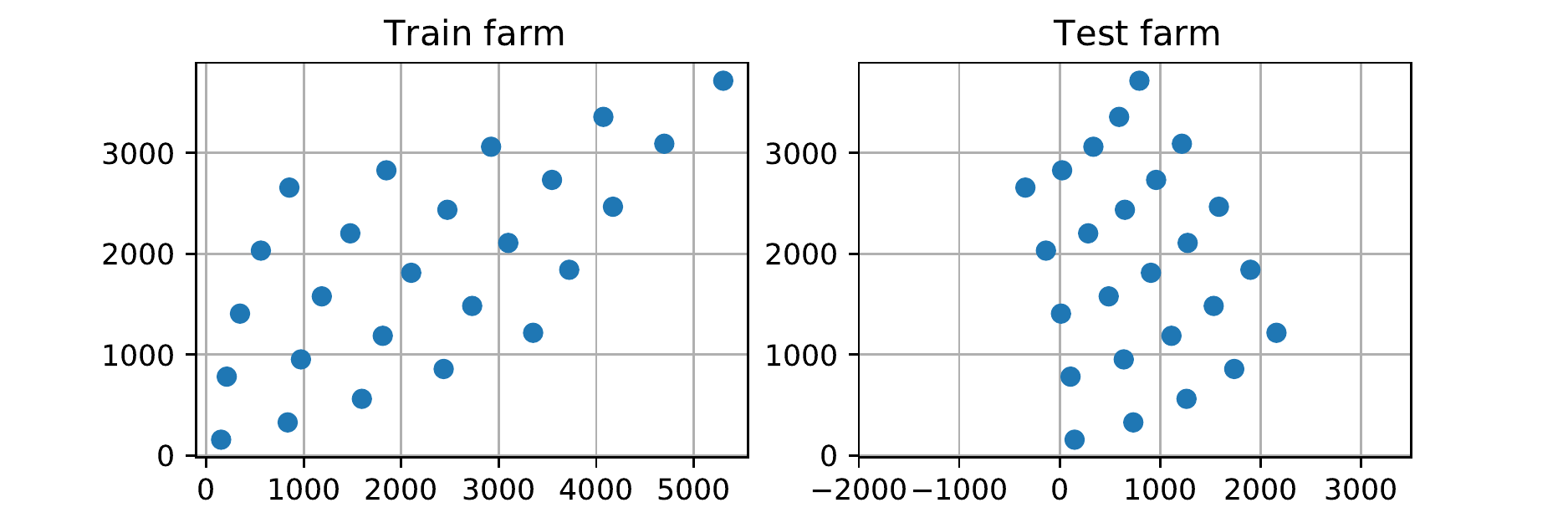}
  \caption{\textbf{Left:} The wind farm creating the training set. \textbf{Right} The wind farm used for the test set.}
  \label{fig:traintestfarm}
\end{figure}

\paragraph{Treatment of angles}
\label{sec:app_windfarms}
As shown in Figure~\ref{fig:relpar}, the angle of the directional vector defined by the sender turbine $i$ to the receiver turbine $j$ $\phi_{ij}$, and distance $d_{ij} $ are used as an edge feature. In real farm monitoring data, the turbine yaw is adjusted by a per-turbine controller during operation according to the wind conditions and may be different for each turbine. The yaw angle $\theta_i$ with respect to north is used as a node feature together with mean and standard deviation of hub-height wind speed when available, and power production. 

\begin{figure}[h!]
  \centering %\fbox{\rule[-.5cm]{0cm}{4cm} \rule[-.5cm]{4cm}{0cm}}
  \includegraphics[width=0.25\textwidth]{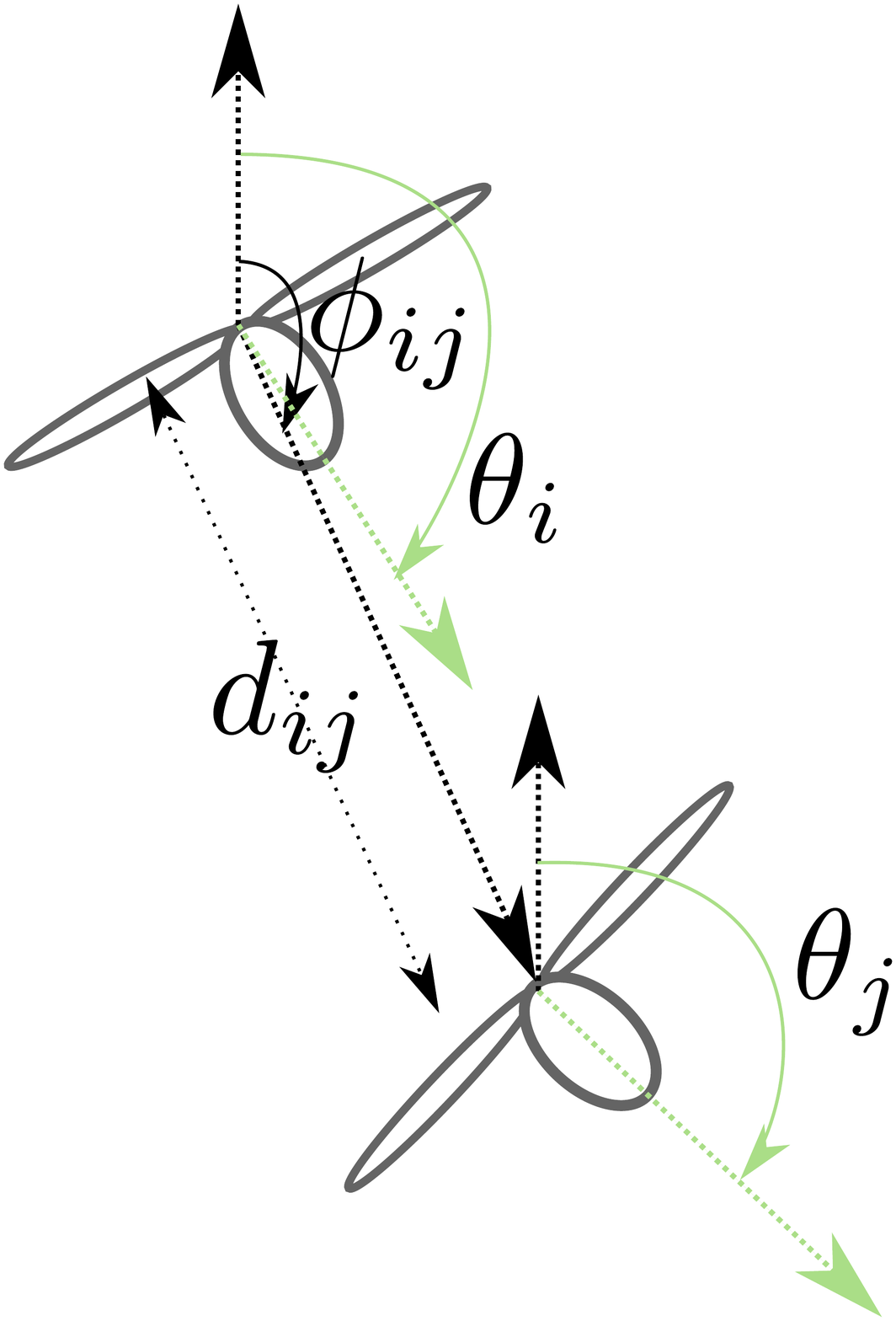}
  \caption{The edge attributes $\E_h$ for edge with $s = i, r = j$ is $(\cos(\phi_{i,j}),\sin(\phi_ij), d_{ij} )$. The nacelle angle $\theta_i$ is included as a node (turbine) attribute.}
  \label{fig:relpar}
\end{figure}

Since the relative angle of the line defined between the turbines and the yaw angles is important, and not the absolute angles, a simple data augmentation procedure is performed during training where both yaw angle and the farm positions are randomly rotated with the same angle. A more general solution to the problem of learning networks that respect transformations to such symmetries
would be to use GNs which are equivariant by design as proposed in \cite{satorras2021n} but this is left for future work.

\paragraph{Further qualitative results on the simulated farm dataset}
In \autoref{fig:rvaesim} some RVAE predictions and actual data from the simulated farm test set are shown.
Note that the configuration of the test farm is unseen and no training is performed.
The RVAE seems to capture the wake-associated wind deficits of the farm in all but one case (second from the right). 

\begin{figure}
  \includegraphics[width=\linewidth]{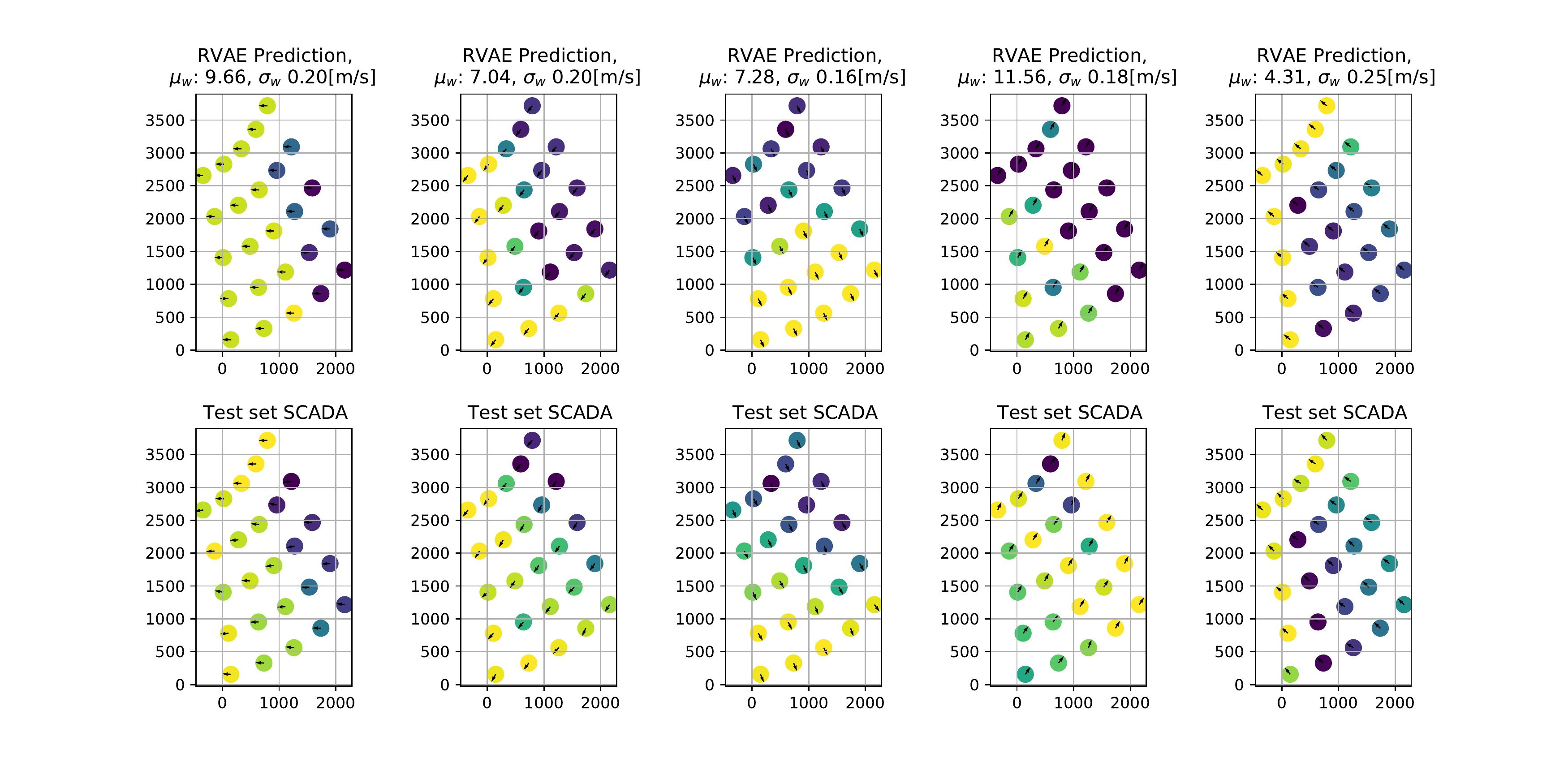}
  \caption{Examples of mean wind from the RVAE when conditioning with the global latent $\mathbf{u}_h$ for an unseen farm (top) and corresponding mean wind from the test dataset (bottom). The arrows correspond to the incoming wind direction, and lighter colors correspond to higher wind speed.}
  \label{fig:rvaesim}
\end{figure}

\subsection{Anholt farm dataset}
\paragraph{Post-hoc model interpretation with gradient sensitivities} In order to identify which nodes are important for the imputed values for a particular turbine, a simple gradient sensitivity technique was employed. 
\label{sensitivities}
For a particular node prediction $\mathbf{v}^T_i \in \V^T$, and a RVAE model 
with recognition model $q_{\bp}$ and generator $p_{\bt}$, we have $\hat{\mathbf{v}}^T_i \sim p_{\bp}(\mathcal{V}_x | G_z ; G_h) q_{\bp}(G_z|G^{C \cup T \setminus b}_x ; G_h)$ where $G^{C \cup T \setminus b}_x$ is the graph that contains the 
masked target nodes and all the edges between context and target nodes.
Since the model is end-to-end differentiable, one can straightforwardly compute an absolute gradient sensitivity as 
\begin{equation}
  I_{\mathbf{v_i}}= \Big| \frac{\partial{\mathbf{\hat{v}}^T_i}}{\partial G^{C \cup T \setminus b}_x} \Big|
\end{equation}
An instance of such maps is visualized for a farm state snapshot and for a representative set of turbines in \autoref{lab:gradsens}. Wind orientation is inferred from turbine nacelle orientation and it is visualized as arrows centered at each turbine and pointing towards the incoming wind. The upwind turbines seem to contribute more to the imputation value of the masked turbines. 
This is aligned with the physics of the problem and in particular with the directionality of the wake effects which is always from upstream to downstream turbines. 
The values of the turbines in the undisturbed boundaries of the farm as in plots indexed 0 and 4 rely on non-upstream neighbors for imputation. This also is expected, as there is no up-wind information for the imputation so the network needs to rely on any neighbors to impute the wind velocity values. 
\begin{figure}[h]
  \includegraphics[width=\textwidth]{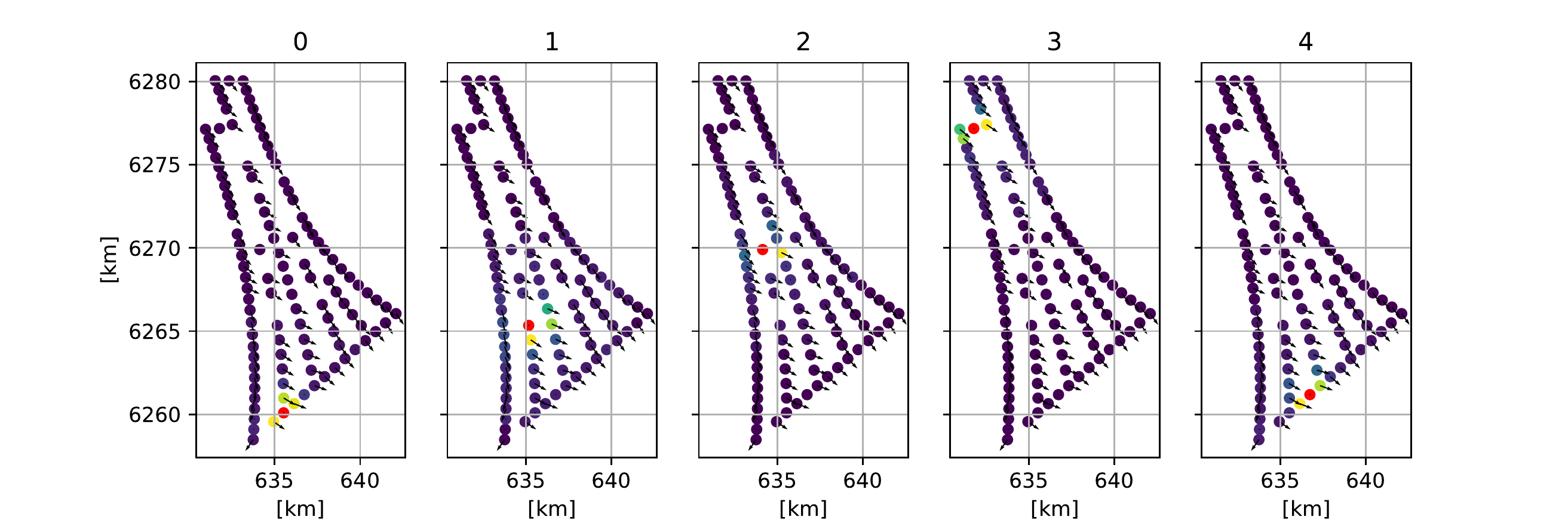}
  \caption{Absolute gradient sensitivity maps for data imputation of SCADA for different turbines, for a single snapshot of the farm conditions. Arrows point against incoming wind. Imputation target turbines in red. The value of neighbors and in particular of upstream neighbors are important.}
  \label{lab:gradsens}
\end{figure}

\section{Supplemental material for the NP section}
\paragraph{Additional details on the training setup}
All experiments were performed using the Adam optimizer \cite{kingma2014adam} with a learning rate of $10^{-4}$ and default parameters. All RVAE models were trained for $4\cdot10^4$ steps and NP models up to $5 \cdot 10^4$ steps with batches of size 16. Each batch contains a random number of context and target points which varies between 3 and 50.
All GN functions involved are feed-forward 3-layer Multi-Layer Perceptron (MLPs) with rectified linear unit non-linearities and no activation in the last layer.
The GN block used is encode-process-decode architecture as in \cite{graphnetpaper} with residual connections. The encoder and decoder contain layers that do not perform message passing but only cast the inputs to a predefined dimension (Graph Independent layers).

\paragraph{Qualitative results on 1D regression meta-learning} In \autoref{fig:meta1drvae} qualitative results for the behavior of the RVAE for varying numbers of context points are shown. In the absence of context points, the model tends to predict the mean which is zero for the training dataset. Upon review, a notebook will be released demonstrating the implementation of the method.%Please refer directly to the notebook for further details on the implementation.

\begin{figure}[h]
\centering
    \includegraphics[width=0.65\textwidth]{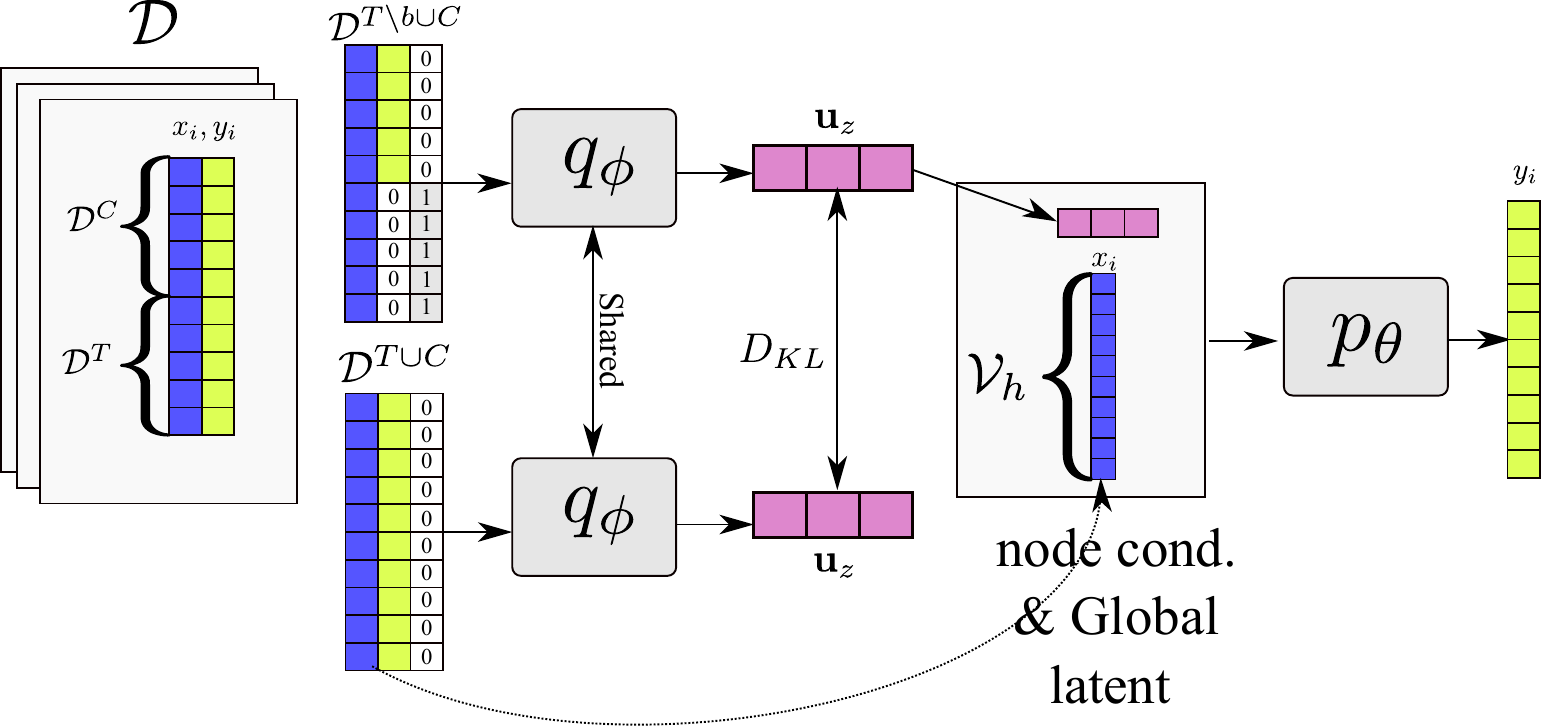}
    \caption{Implementation of NP as a conditional RVAE. The $q_\phi$ is implemented as a deep set, and $p_\theta$ is a node block. The node block is a function that contains a node update function which uses the node inputs and the global context (here the $\mathbf{u}^z$ variable).}
    \label{fig:schemrvaeac}
\end{figure}

 \begin{figure}[h]
   \centering
   \includegraphics[width=0.65\textwidth]{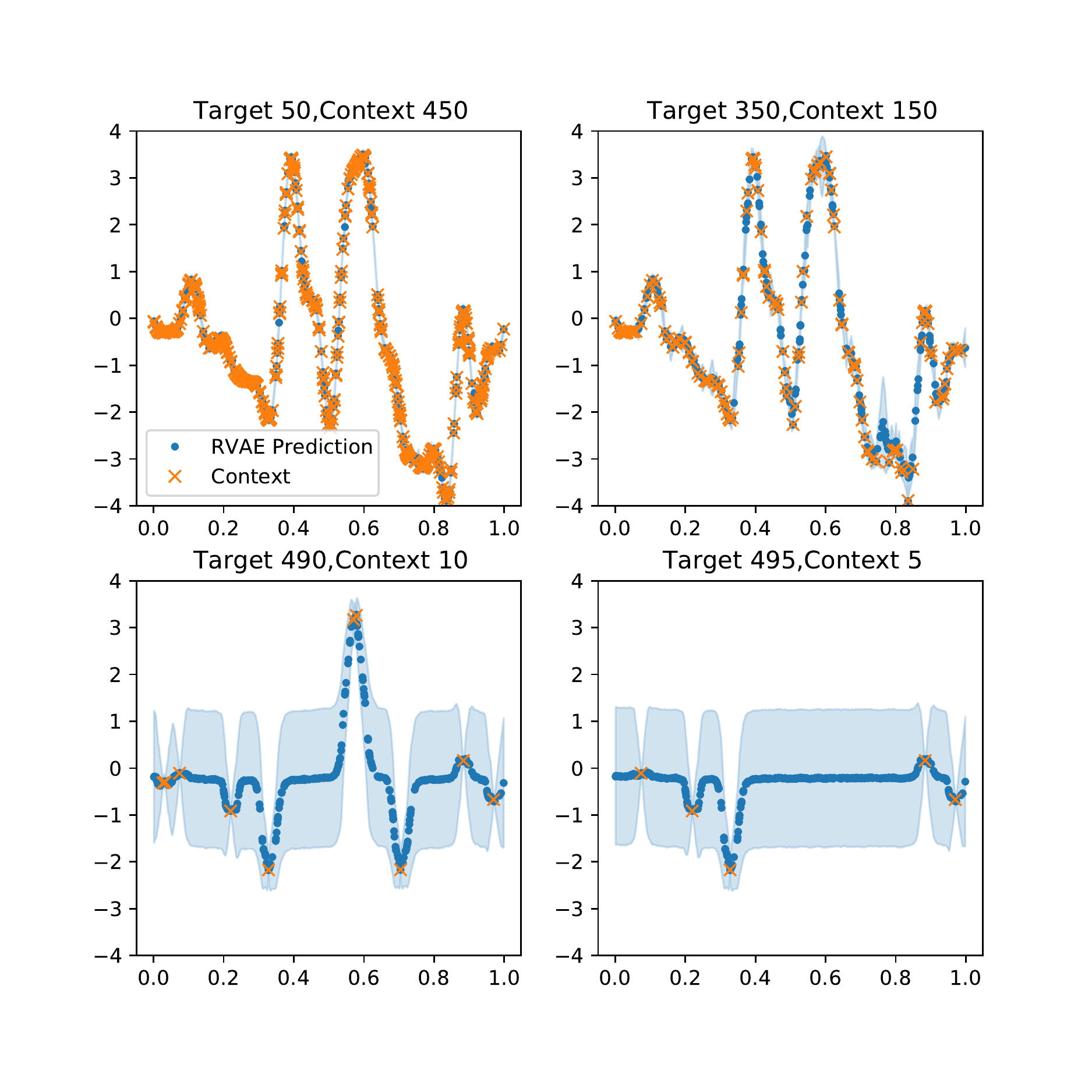}
   \caption{Qualitative results for 1D regression meta-learning with an RVAE. The model was trained using both node and edge features and with samples from GPs that had the same parameters.}
   \label{fig:meta1drvae}
 \end{figure}

 \subsection{Implementation of NP as a VAE with arbitrary conditioning}
  In order to make the exposition of the main text clearer in what follows the ELBO objective for a RVAE model with arbitrary conditioning and a DeepSet encoder is detailed. 
 In the NP implementation a set of context $\mathcal{D}^C$ and target $\mathcal{D}^T$ points are encoded through $q_{\bp}$ to a global Gaussian latent variable $\mathbf{u_z}$. Now consider the NP problem as a missing data imputation problem, where the context points, $\mathcal{D}^{C} : \{x_C, y_C\}$ are observed sets of points from a function, and the target points $\mathcal{D}^{T} : \{ x_T,y_T\}$ are points coming from the same function where we only have their $x$ coordinate ($x_T$) and we seek to evaluate the $y_T$ values. The NP approach to this problem is to separately encode  $\mathcal{D}^{C \cup T}$ and $\mathcal{D}^{C}$ to a Gaussian latent space $\mathbf{u}_z$ using a DeepSet encoder, use the known $x_T$ as conditioning, and compute $y_T$ using $\mathbf{u}_z$ and $x_T$ (and optionally a deterministic $r$ introduced in \cite{kim2019attentive}). The loss function for NPs given in \autoref{eq:nploss} can be cast as an arbitrary conditioning objective as in \cite{ivanov2018variational}.

 The VAE-AC\cite{ivanov2018variational} approach to an imputation problem such as this one, is to augment the dataset with an additional variable $b$ which signifies whether 
 points are observed or unobserved. The $D_{KL}$ term of the ELBO objective drives the latent representation $\mathbf{u}_z$ of the recognition network to be similar when observing $\mathcal{D}^{C}$ and 
 $\mathcal{D}^{C \cup T}$. A network with parameters $\bt$ is jointly trained to  yield $y_T \sim p_{\bt}(y_T | \mathbf{u}_z, x_T)$. 
 The arbitrary conditioning ELBO objective in that case reads,
 \newcommand{\DCTb}{\mathcal{D}^{C \cup T \setminus b}}
 \newcommand{\DCT}{\mathcal{D}^{C \cup T}}
\begin{align}
  \label{eq:nploss}
  \log p(y_T|x_T,x_C,y_C) \geq & \mathbb{E}_{q_{\phi}(z | \DCT)}\Big[ \sum_{i \in T} \log p_\theta(y_i|z, x_i) + \log \frac{q(z|\DCTb)}{q(z|\DCT)} \Big] \nonumber \\
                          %& \mathbb{E}_{q_{\phi}(z | \mathcal{D}^{C \cup T})}\Big[ \sum_{i \in T} \log p_\theta (y_i|z, x_i) \Big] - \mathbb{E}_{q_{\phi}(z | \mathcal{D}^{C \cup T})}\Big[ \log \frac{q(z|\mathcal{D}^{C\cup T})}{q(z|\mathcal{D}^{C})} \Big]  \nonumber  \\
                          & \mathbb{E}_{q_{\phi}(z | \DCT)}\Big[ \sum_{i \in T} \log p_\theta(y_i|z, x_i) \Big] - D_{KL}(q_{\phi}(z | \DCT) || q_{\phi}(z | \DCTb)) 
\end{align}
The VAE-AC approach is particularly convenient for the RVAE which contains message passing layers, since no special message passing needs to be performed between observed and unobserved points. In \autoref{fig:schemrvaeac} a more detailed computational diagram is shown.

\end{document}